\documentclass[12pt]{iopart}
\usepackage{graphicx}
\usepackage{iopams}
\usepackage[pdfborder=000, colorlinks, linkcolor=blue, citecolor=blue]{hyperref}
\usepackage[numbers,sort&compress]{natbib}
\usepackage{color}

\usepackage{caption}
\begin{document}

\title{Multilayer social reinforcement induces bistability on multiplex networks}

\author{Longzhao Liu$^{1,2,4,6}$, Xin Wang$^{1,3,6}$, Shaoting Tang$^{1,6,7}$, Hongwei Zheng$^{5,7}$ and Zhiming Zheng$^{1,5,6}$}

\address{$^1$ LMIB, NLSDE, BDBC, School of Mathematical Sciences, Beihang University, Beijing 100191, China}
\address{$^2$ ShenYuan Honor School, Beihang University, Beijing 100191, China}
\address{$^3$ Department of Mathematics, Dartmouth College, Hanover, NH 03755, USA}
\address{$^4$ Northwestern Institute on Complex Systems, Northwestern University, Evanston, IL 60208, USA}
\address{$^5$ Institute of Artificial Intelligence and Blockchain, Guangzhou university, Guangdong province 510006, China}
\address{$^6$ PengCheng Laboratory, Shenzhen, 518055 , China}
\address{$^7$ Author to whom any correspondence should be addressed}

\ead{tangshaoting@buaa.edu.cn, hwzheng@pku.edu.cn}

\vspace{10pt}

\begin{abstract}
Social reinforcement mechanism, which characterizes the promoting effects when exposing to multiple sources in social contagion process, is ubiquitous in information-technology ecosystem and has aroused great attention in recent years. While the impacts of social reinforcement on single-layer networks are well-documented, extension to multilayer networks is needed to study how reinforcement from different social circles influences the spreading dynamics. To this end, we incorporate multilayer social reinforcement into ignorant-spreader-ignorant (SIS) model on multiplex networks. Our theoretical analysis combines pairwise method and mean-field theory and agrees well with large-scale simulations. Surprisingly, we find this complex social contagion mechanism triggers the emergence of bistability phenomena, where extinction and outbreak states coexist. In particular, the hysteresis loop of stationary prevalence occurs in this bistable region, explaining why the fight towards the spread of rumors is protracted and difficult in modern society. Further, we show that the final state of bistable regions depends on the initial density of adopters, the critical value of which decreases as the contagion transmissibility or the multilayer reinforcement increases. In particular, we highlight two possible conditions for the outbreak of social contagion: to possess large contagion transmissibility, or to possess large initial density of adopters with strong multilayer reinforcement. Our results unveil the non-negligible power of social reinforcement on multiplex networks, which sheds lights on designing efficient strategies in spreading behaviors such as marketing and promoting innovations.
\end{abstract}


\section{Introduction}
Social contagion describes a variety of behavioral imitations caused by social influence and is particularly ubiquitous in the digital age \cite{axelrod1997dissemination, del2016spreading, stewart2019information}. In order to predict and control the collective contagion phenomena on large-scale social networks, such as the spread of rumors, social norms and online behaviors, scientists have made great efforts on understanding the underlying dynamical mechanisms \cite{zhang2016dynamics, castellano2009statistical, pei2013spreading, liu2019homogeneity}. Early studies argued that the dynamical processes of social contagion are similar to disease spreading in the sense that a simple contact with a single infected individual can trigger diffusion \cite{daley1964epidemics,li2014rumor,vega2017rumor}. Therefore, many epidemic-like models are proposed, which are called simple contagion models. These models provide profound insights into many physical phenomena on social networks, such as the cascading process of information diffusion, the co-contagion dynamics on multiplex networks, and etc \cite{wang2017promoting,moreno2004dynamics}.

However, simple contagion model can not deal with the spread of more complex social behaviors, especially when the behaviors are risky, costly or polarized, ranging from the spread of public health behaviors (e.g., vaccine, vaping, diet) to social movement \cite{centola2010spread,wang2020public,fu2017dueling}. For instance, a well-known complex mechanism of social contagion is called the {\it social reinforcement}. It corresponds to the fact that exposure to multiple sources with a same stimuli would give individuals more confidence to participate than multiple exposures to the same source, leading to a significant promotion on social transmission \cite{romero2011differences,centola2007cascade}. As McAdam and Paulsen concluded \cite{mcadam1993specifying}, ``the fact that we are embedded in many relationships means that any major decision we are contemplating will likely be mediated by a significant subset of those relationships". Nevertheless, this empirically confirmed phenomenon is fundamentally different from the view of simple contagion model \cite{hodas2014simple,aral2017exercise}. Hence, more complex mechanisms were further incorporated into dynamical models. These complex contagion models lead to abundant intriguing findings, some of which even conflict with conclusions of simple contagion models \cite{granovetter1977strength,melnik2013multi}. For example, threshold models assumed that social contagion only happens when the number (or fraction) of exposures to multiple sources exceeds a given threshold and suggested the dependence of diffusion results on initial density of adopters \cite{watts2002simple}. In addition, Centola {\it et al.} proved that the clustering network structure suppressed simple contagion while surprisingly facilitated the spread of behaviors that require social reinforcement \cite{centola2007complex}. Another framework modified epidemic-like models by increasing transmission rate when exposing to multiple sources, and showed the promoting effect of social reinforcement on contagion processes \cite{zheng2013spreading,neuhauser2020multibody}. Another well-known complex mechanism is caused by the interplay between spreading dynamics and network topology. For example, Gross {\it et al.} studied contagion process on adaptive networks where the connection between nodes relies on their states, and showed that this dynamics-topology interaction could lead to the emergence of bistable region where healthy and endemic states co-exist \cite{gross2006epidemic}. Besides, Iacopini {\it et al.} noted that the existence of high-order structures embedded on single-layer networks, such as full triangles, could also induce bistable phenomena \cite{iacopini2019simplicial}.

Moreover, complex mechanisms depicting interactions between multiple contagion processes on networked systems, such as the spreading of interacting diseases, were widely investigated \cite{wang2019coevolution}. Chen {\it et al.} explored cooperative contagion processes where individuals infected by one disease are more susceptible to the other, and showed that large degree of cooperation could lead to the occurrence of abrupt phase transition and bistability \cite{chen2017fundamental}.  Soriano-Pa{\~n}os {\it et al.} described the interaction between cooperative or competitive diseases by increasing or decreasing the susceptibility of individuals infected by one of the diseases, and also found bistability caused by cooperativity between diseases \cite{soriano2019markovian}.  Sanz {\it et al.} proposed a general framework analyzing simultaneous spreading of two interacting diseases and derived complex phase diagrams \cite{sanz2014dynamics}. Pinotti {\it et al.} explored a three-player pathogen system where competition and cooperation coexist, and showed that the presence of cooperative pathogen could lead to some intriguing phenomena including non-monotonic boundaries separating phase diagram and bistability \cite{pinotti2020interplay}. There were also many works extending the framework from single-layer networks to multilayer networks \cite{brodka2020interacting}. For example, Wu {\it et al.} studied discrete-time Markov-Chain model depicting the spreading of two diseases in multiplex networks and derived epidemic thresholds \cite{wu2020spreading}.

Recently, owing to the development of various social medias, many social behaviors spread in multiple social circles rather than in a single social circle, which are naturally modeled as multilayer networks \cite{salehi2015spreading,soriano2019explosive,de2016physics,bianconi2017epidemic}. Each layer represents a single social circle which could be an online social platform or a network of offline relationships, such as friendships and colleagues. Similar to social reinforcement in a single-layer network, there exists significant difference between exposure to multiple social circles (multilayer reinforcement) and multiple exposures to the same social circle (intra-layer reinforcement)\cite{chen2018complex}. Specifically, individuals would be more convinced and have higher possibility for diffusion when receiving the same informative stimulus from different social circles. However, previous studies mainly concentrated on simple interactions between nodes and its replicas in different layers \cite{cozzo2013contact,gomez2013diffusion}. The detailed impacts of multilayer reinforcement mechanism that engineers complex social contagion remain largely unknown.

To fill this gap, here we propose a theoretical framework that incorporates multilayer reinforcement into ignorant-spreader-ignorant (SIS) model to study the spreading dynamics on multiplex networks. We find this complex social contagion mechanism not only expands dissemination, but also results in the emergence of bistable region, within which extinction and outbreak states coexist. Within bistable region, the final state depends on the initial density of adopters.  We also detailedly discuss the hysteresis loop and the unstable equilibrium manifolds occurred in the bistable region. Through observing phase diagrams, we highlight two conditions for the outbreak of social contagion: (i) to possess large transmissibility; (ii) to possess large initial density of adopters with strong multilayer reinforcement. As the second condition cannot be obtained by simple contagion model and is less intuitive, our results stress the unneglectable role of complex social contagion and are in line with the findings of previous experimental studies that critical masses are required for establishing collective behaviors such as social changes \cite{centola2018experimental}. In addition, similar phenomena are also observed on finite-size heterogeneous multiplex networks. Our findings provide valuable insights toward dynamical evolutions of complex social contagion on multiplex networks, which are of vital importance for understanding collective online behaviors in the era of social media. In particular, the emerging bistable phenomenon on multiplex
networks indicates that the information (truth, ideas, advertisements, etc.) can stride across
the trap of extinction and become widespread by selecting a large proportion of initial
spreaders on multiple platforms, which implies effective strategies for controlling rumors,
promoting innovations and marketing.  \cite{monsted2017evidence,mahajan2010innovation,krapivsky2011reinforcement}.

\section{Model}\label{model}
\begin{figure}
\center
\includegraphics[width=12cm]{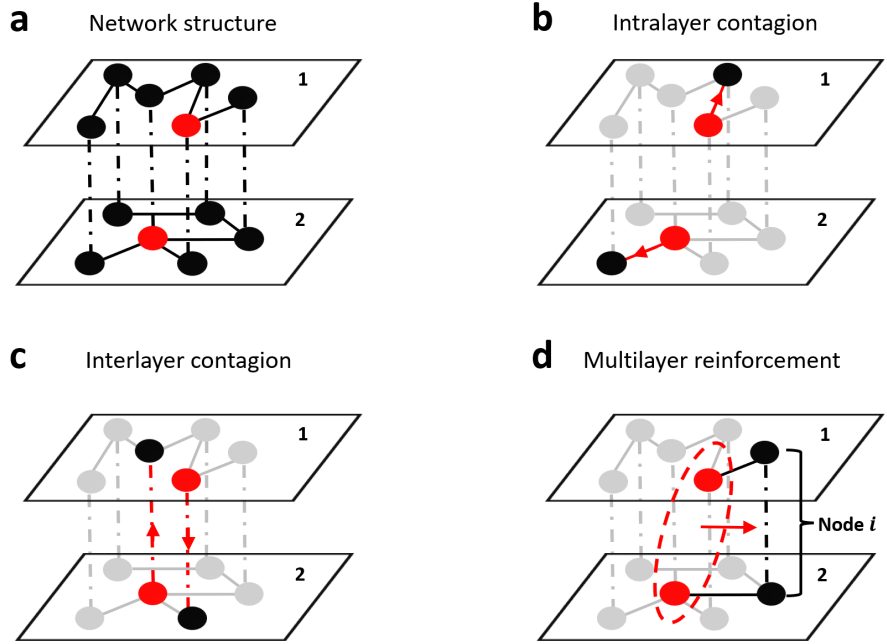}

\caption{Schematic of modeling framework.  (a) Network structure. The underlying network is composed of two layers with the same nodes, where interlayer edges only connect entities and their replicas. Node's state is represented by different colors: spreader (red), ignorant (black). (b)-(d) Dynamical model. Our model includes three dynamical processes, whose propagation path is highlighted by red arrows in each subfigure. (b) Intralayer contagion. In each layer, an ignorant has a probability to be infected by its spreader neighbors. (c) Interlayer contagion. An ignorant may change its state to a spreader if its replica is a spreader. (d)  Multilayer reinforcement. When node $i$ is connected to at least one spreader in each layer and is able to receive stimuli from both platforms at the same time, $i$ randomly chooses a layer and becomes a spreader with a certain probability.}
\label{Dynamical model}
\end{figure}

Consider an undirected multiplex network with two layers, denoted as layer 1 and layer 2. Each layer stands for a social circle, which is an online social platform or an offline relationship network composed of interacting individuals such as friends, families and colleges. Both layers have the same nodes and interlayer edges only connect entities with their replicas, as shown in figure  \ref{Dynamical model}(a).

Here we stress that our main purpose is to explore the influence of complex social contagion aroused by multiple social circles, i.e., multilayer reinforcement mechanism. Therefore, we adopt ignorant-spreader-ignorant (SIS) model rather than complex contagion model to characterize the intralayer spreading process, which provides better analytical insights while at expense of being less realistic. In this situation, spreader ($S$) represents individuals who adopt norms (information, cognition, attitudes, behaviors and so on) and are willing to spread, while ignorant ($I$) stands for nodes who do not adopt norms or have no motivation to spread, corresponding to infected and susceptible state in epidemiology, respectively. In each layer $k$  $(k=1,2)$, spreader has a probability $\lambda_k$ to spread norms to its ignorant neighbors and becomes ignorant with probability $\mu_k$, as shown in figure  \ref{Dynamical model}(b).

Beyond intralayer contagion processes, norms also diffuse across layers, i.e., the interactions between layers. First, complying with previous studies, we utilize interlayer contagion processes where ignorant has a probability $p$ to become a spreader once its counterpart is in $S$ state, which depicts the interaction between agents and their replicas, as shown in figure  \ref{Dynamical model}(c) \cite{li2015multiple}. Then, we introduce multilayer reinforcement mechanism to mimic the fact that exposure to multiple social circles is more convincing than multiple exposures to a single social circle. At each time step, if an individual can receive norms from both social circles, i.e. he/she has at least one spreader neighbor in each layer, then with probability $\gamma$ the individual would randomly choose one layer to spread the norms (figure  \ref{Dynamical model}(d)).  The parameter $\gamma$ reflects the strength of multilayer reinforcement.

In summary, our model is composed of three processes: intralayer contagion, interlayer contagion and multilayer reinforcement, which is as follows:
\begin{enumerate}
	\item\textbf{Intralayer contagion.} At each time step, in layer $k$, $S$ makes its neighbors in $I$ state become $S$ with probability $\lambda_k$. Meanwhile, $S$ turns into $I$ with probability $\mu_k$.
   \item\textbf{Interlayer contagion.} At each time step, $I$ becomes $S$ with probability $p$ if its replica is a spreader.
   \item\textbf{Multilayer reinforcement.} At each time step, a node has a probability $\gamma$ to randomly choose a layer and become a spreader if there exists at least one spreader in its neighborhood of each layer.
\end{enumerate}

\section{Theoretical framework}

In this section, we explore dynamical equations of our model on homogeneous multiplex systems, composed of two Erd\"{o}s-R\'{e}nyi (ER) networks with average degree $\langle k_1\rangle$ and $\langle k_2\rangle$. The major challenge is the dynamical correlation between two layers caused by interlayer contagion and multilayer reinforcement. Specifically, the counterpart of a spreader is more likely to be a spreader than the counterpart of an ignorant. Classic Mean-field approximation does not consider the dynamical correlation, which leads to the deviation of theoretical predictions from Monte Carlo simulations (see \ref{MFA}). To overcome the difficulty, we conduct theoretical analysis by combining pairwise method and mean-field theory.

Here, we define $\rho_{X,Y}(t)$ as the probability that the individual is $X$ state in layer 1 and $Y$ state in layer 2. Clearly, each individual has four possible states and we have $\rho_{S,S}(t)+\rho_{S,I}(t)+\rho_{I,S}(t)+\rho_{I,I}(t)=1$. To begin, we explore intralayer contagion process. Let $\theta_k(t)$ represent the probability that a node in layer $k$ is not convinced by its neighbors of the same layer. Utilizing the mean-field theory, $\theta_k(t)$ can be approximated by

\begin{equation}
\eqalign{
\theta_1(t)&=(1-\lambda_1(\rho_{S,S}(t)+\rho_{S,I}(t)))^{\langle k_1\rangle}\\
\theta_2(t)&=(1-\lambda_2(\rho_{S,S}(t)+\rho_{I,S}(t)))^{\langle k_2\rangle}.
}
\label{1}
\end{equation}

Then, we examine the influence of multilayer reinforcement. Denote $\delta_k(t)$ as the probability that the multilayer reinforcement does not make ignorant turn into a spreader in layer $k$. We have $\delta_1(t)=\delta_2(t)=\delta(t)$, which can be approximated by

\begin{equation}
\eqalign{
\delta(t)=1-0.5&*\gamma\\
&*(1-(1-\rho_{S,S}(t)-\rho_{S,I}(t))^{\langle k_1\rangle})\\
&*(1-(1-\rho_{S,S}(t)-\rho_{I,S}(t))^{\langle k_2\rangle}),
}
\label{2}
\end{equation}
where $(1-(1-\rho_{S,S}(t)-\rho_{S,I}(t))^{\langle k_1\rangle})$ and $(1-(1-\rho_{S,S}(t)-\rho_{I,S}(t))^{\langle k_2\rangle})$ account for the probabilities that there exists at least one spreader neighbor in layer 1 and in layer 2, respectively.

Furthermore, let $g^I_k(t)$ and $g_k^S(t)$ represent the probabilities that ignorant in layer $k$ is not convinced if its counterpart is in ignorant and spreader state, respectively. The probabilities describe the integrated influence of intralayer contagion, interlayer contagion and multilayer reinforcement. Because the three processes are approximately independent in our model, the probabilities read

\begin{equation}
\eqalign{
g^I_k(t)&=\theta_k(t)*\delta(t)\\
g^S_k(t)&=\theta_k(t)*\delta(t)*(1-p).
}
\label{3}
\end{equation}

Finally, the temporal evolutionary equations of $\rho_{S,S}(t)$ can be written as
\begin{equation}
\eqalign{
\frac{d\rho_{S,S}}{dt}&=-\rho_{S,S}\{1-(1-\mu_1)(1-\mu_2)\}+\rho_{S,I}(1-\mu_1)(1-g^S_2)\\
&+\rho_{I,S}(1-\mu_2)(1-g^S_1)+\rho_{I,I}(1-g^I_1)(1-g^I_2),\\
}
\label{9}
\end{equation}
where the first term represents the outflow from the $S-S$ class and the last three terms stand for transition from the other states to $S-S$ state. Similarly, we can derive the dynamical evolutions of $\rho_{S,I}(t)$, $\rho_{I,S}(t)$ and $\rho_{I,I}(t)$. For the sake of theoretical analysis and readability, $\rho_{I,I}(t)$ is automatically substituted for $(1-\rho_{S,S}(t)-\rho_{S,I}(t)-\rho_{I,S}(t))$. Thus, the evolutionary equations of our model read as follows:

\begin{equation}
\eqalign{
\frac{d\rho_{S,S}}{dt}&=-\rho_{S,S}\{1-(1-\mu_1)(1-\mu_2)\}+\rho_{S,I}(1-\mu_1)(1-g^S_2)\\
&+\rho_{I,S}(1-\mu_2)(1-g^S_1)+(1-\rho_{S,S}-\rho_{S,I}-\rho_{I,S})(1-g^I_1)(1-g^I_2)\\
\frac{d\rho_{S,I}}{dt}&=\rho_{S,S}(1-\mu_1)\mu_2-\rho_{S,I}\{1-(1-\mu_1)g^S_2\}\\
&+\rho_{I,S}(1-g^S_1)\mu_2+(1-\rho_{S,S}-\rho_{S,I}-\rho_{I,S})(1-g^I_1)g^I_2\\
\frac{d\rho_{I,S}}{dt}&=\rho_{S,S}(1-\mu_2)\mu_1+\rho_{S,I}(1-g^S_2)\mu_1\\
&-\rho_{I,S}\{1-(1-\mu_2)g^S_1\}+(1-\rho_{S,S}-\rho_{S,I}-\rho_{I,S})g^I_1(1-g^I_2).\\
}
\label{4}
\end{equation}

Equation  (\ref{4}) can be simply written as
\begin{equation}
\frac{d\boldsymbol\rho}{dt}=\boldsymbol f(\boldsymbol\rho),
\label{5}
\end{equation}
where $\boldsymbol{\rho}(t)=(\rho_{S,S}(t),\rho_{S,I}(t),\rho_{I,S}(t))$. This indicates that the dynamical system is 3-dimension autonomous. Thus, the stability of the fixed points directly determines evolutionary results of the system. We define $\boldsymbol\rho_f$ as the fixed point of equation  (\ref{4}), which satisfies
\begin{equation}
\boldsymbol f(\boldsymbol\rho_f)=\boldsymbol 0.\\
\label{6}
\end{equation}
In particular, $\boldsymbol\rho_f$ represents the final state of the system if and only if the fixed point is stable, i.e.,
\begin{equation}
\Lambda_{max}(\frac{d\boldsymbol f}{d\boldsymbol \rho}|_{\boldsymbol \rho=\boldsymbol \rho_f})<0,\\
\label{7}
\end{equation}
where $\Lambda_{max}(J)$ is the largest eigenvalue of matrix $J$.

We notice that there is at least one fixed point ($\boldsymbol \rho_f=\boldsymbol 0$), which represents that all individuals are in $I-I$ state. When the initial density of spreaders is small, norms go extinct if and only if $\boldsymbol \rho_f=\boldsymbol 0$ is stable, i.e.,

\begin{equation}
\Lambda_{max}(\frac{d\boldsymbol f}{d\boldsymbol \rho}|_{\boldsymbol \rho=\boldsymbol 0})<0.\\
\label{8}
\end{equation}

\section{Results}
\subsection{Complex social contagion on homogeneous multiplex networks}
\begin{figure}
\center
\includegraphics[width=12cm]{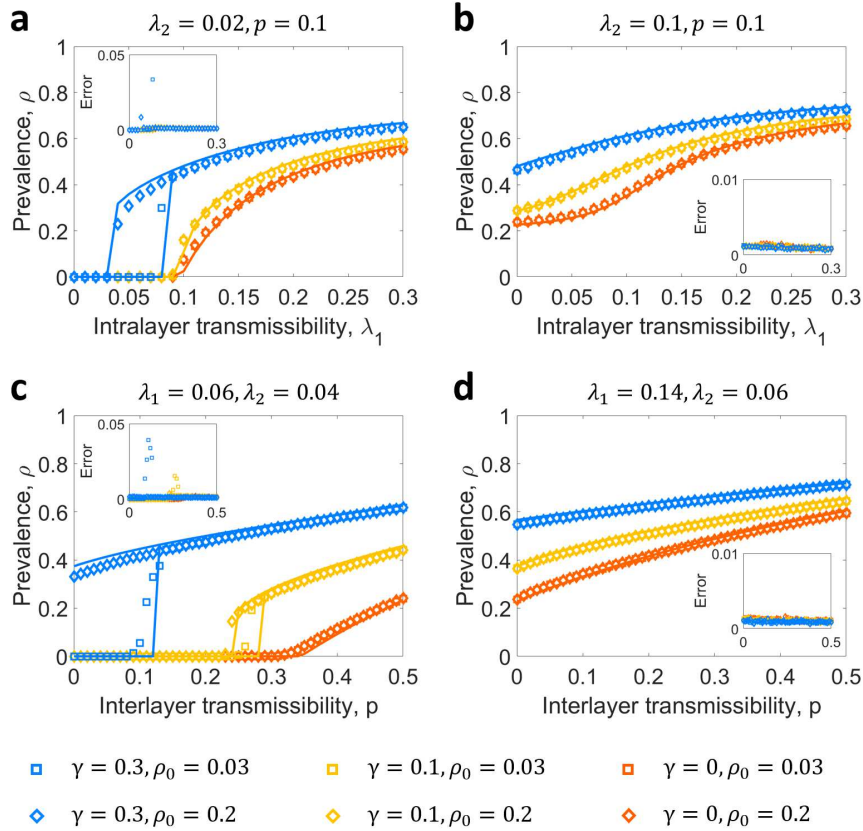}

\caption{Emergence of bistable phenomena. Prevalence curves are shown against (a)-(b) intralayer transmissibility and (c)-(d) interlayer transmissibility for different values of multilayer reinforcement and initial density of adopters. In each figure, $\gamma=0$ (red), $\gamma=0.1$ (yellow), $\gamma=0.3$ (blue) and $\rho_0=0.03$ (square), $\rho_0=0.2$ (diamond).  Simulation results are averaged over 30 independent runs, whose standard errors are shown in subplots. In most cases, errors are within 0.01. However, errors can exceed 0.01 when the system is close to the critical points, as the stochastic fluctuation of simulations is relatively large around the thresholds. Theoretical predictions solved by equation  (\ref{4}) are shown by solid lines. Note that for $\gamma=0.3$ in (a) and $\gamma=0.1$ or $\gamma=0.3$ in (c), there emerges bistable region where outbreak and extinction states co-exist. Parameters: $\mu_1=0.6$, $\mu_2=0.6$. In addition, (a) $\lambda_2=0.02$, $p=0.1$, (b) $\lambda_2=0.1$, $p=0.1$, (c) $\lambda_1=0.06$, $\lambda_2=0.04$, and (d) $\lambda_1=0.14$, $\lambda_2=0.06$.}
\label{intralayer}
\end{figure}
In this section, we explore how our model behaves on homogeneous multiplex networks, especially the influence of multilayer reinforcement on the prevalence of norms and the critical properties that separate outbreak and extinction. We start from homogeneous multiplex networks composed of two Erd\"{o}s-R\'{e}nyi (ER) graphs with $N=10000$ nodes, the average degree of which are $\langle k_1\rangle=6$ and $\langle k_2\rangle=8$, respectively. Initially, we randomly set a certain fraction of population as adopters, denoted by $\rho_0$, who are spreaders in both layers\cite{wang2018social}. Here we run 300-step simulations to ensure that the system has reached the dynamic equilibrium.

\begin{figure}
\center
\includegraphics[width=12cm]{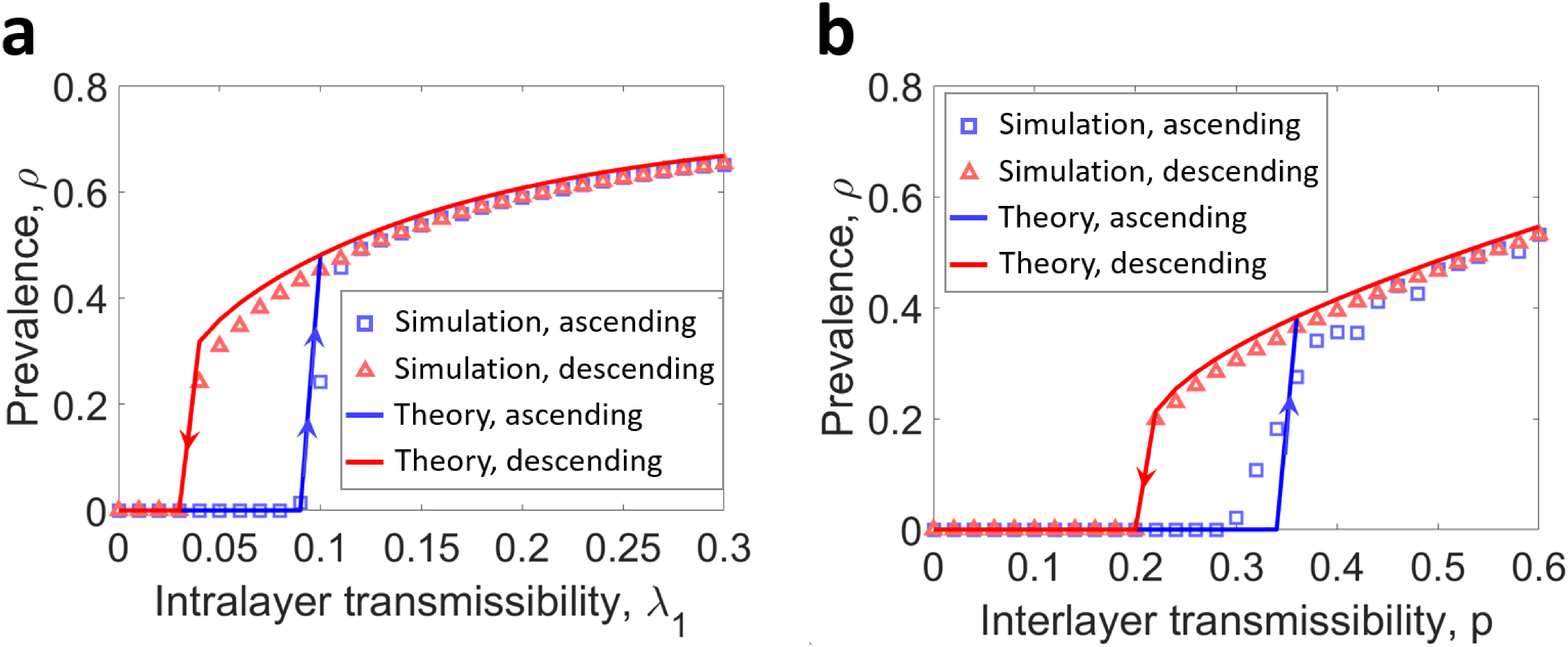}

\caption{Hysteresis phenomena. Shown are the hysteresis loops of the stationary prevalence with respect to changes in (a) intralayer transmissibility and (b) interlayer transmissibility. Two different evolutionary routes arise: prevalence increases from zero to positive values along the ascending path (blue squares) with $\lambda_1$ or $p$ grows, while recovers to zero along the descending path (red triangles) as transmissibility decreases. Theoretical predictions (solid lines) are in good agreement with simulations in most situations. The relatively large stochastic fluctuation of simulations only occurs around the thresholds. Parameters: (a) $\lambda_2=0.02$,$p=0.1$,$\gamma=0.3$,$\mu_1=\mu_2=0.6$; (b) $\lambda_1=0.06$, $\lambda_2=0.04$, $\gamma=0.12$, $\mu_1=\mu_2=0.6$.}
\label{hysteresis}
\end{figure}

\begin{figure}
\center
\includegraphics[width=12cm]{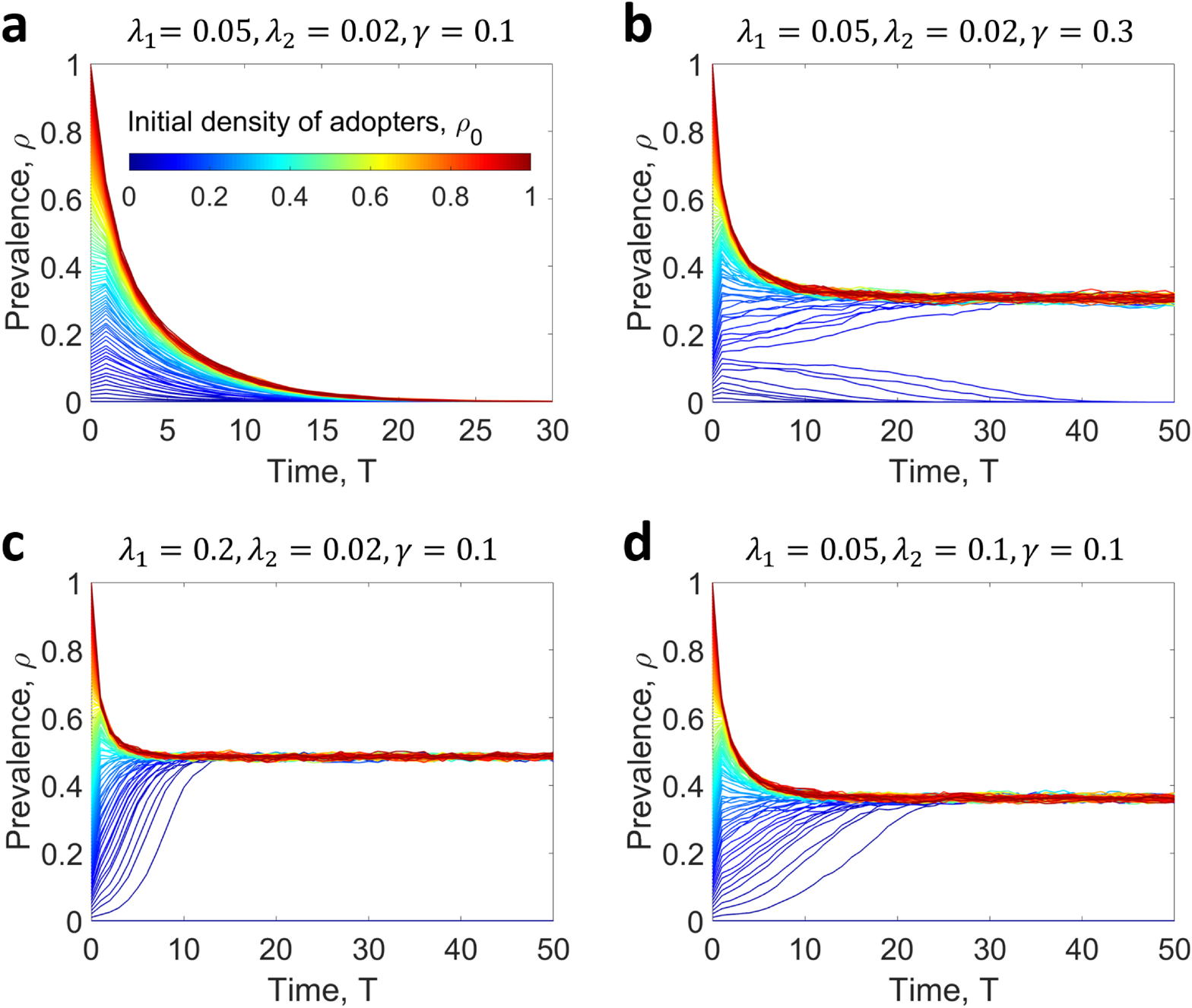}

\caption{Effect of initial density of adopters. (a)-(d) Time evolutions for the prevalence are presented under different combinations of transmission parameters. In each figure, a single curve corresponds to one value of different initial densities of adopters. (b) We find that when $\gamma$ is relatively large, the initial condition determines the final state of bistable region, either outbreak or extinction. Parameters: $\mu_1=0.6$, $\mu_2=0.6$, $p=0.1$. In addition, (a) $\lambda_1=0.05$, $\lambda_2=0.02$, $\gamma=0.1$, (b) $\lambda_1=0.05$, $\lambda_2=0.02$, $\gamma=0.3$, (c) $\lambda_1=0.2$, $\lambda_2=0.02$, $\gamma=0.1$, and (d) $\lambda_1=0.05$, $\lambda_2=0.1$, $\gamma=0.1$. }
\label{Time evolutions}
\end{figure}

To begin with, we show prevalence curves as a function of intralayer transmissibility (figure  \ref{intralayer}(a) and figure  \ref{intralayer}(b)) and interlayer transmissibility (figure  \ref{intralayer}(c) and figure  \ref{intralayer}(d)) under different combinations of multilayer reinforcement and initial density of adopters. All subfigures verify the intuition that multilayer reinforcement promotes social contagion and show that our theoretical predictions agree well with simulation results. The case $\gamma=0$ (the red curves) is equivalent to SIS model on multiplex networks , which displays continuous phase transition \cite{cozzo2013contact}. Nevertheless, the case $\gamma=0.3$ (the blue curves) shows large differences in dependence of prevalence on intralayer transmissibility (figure  \ref{intralayer}(a)) and interlayer transmissibility(figure  \ref{intralayer}(c)). Phase transition appears at lower value of $\lambda_1$ or $p$, and becomes discontinuous. Another interesting phenomenon is the emergence of bistable region, where outbreak and extinction states coexist. Specifically, for $\lambda_1\in(0.03,0.08)$ in figure \ref{intralayer}(a) and $p\in(0,0.12)$ in figure \ref{intralayer}(c), norms outbreak if $\rho_0=0.2$, while go extinct if $\rho_0=0.03$. It indicates that the final state in bistable region might depend on the initial density of adopters. Owing to the spontaneous recovery mechanism ($\mu_1$ and $\mu_2$), the system would finally reach a dynamic equilibrium. We conduct additional simulations under different $\mu_1$ and $\mu_2$, and obtain similar bistable phenomena and discontinuous phase transitions even under weak multilayer reinforcement when $\mu_1$ and $\mu_2$ are small (see figure \ref{mu} in \ref{appendix}).

In addition, figure \ref{hysteresis} shows the emergence of hysteresis loops of stationary prevalence with respect to changes in intralayer transmissibility and interlayer transmissibility. Here two evolutionary routes can be observed.  One is the ascending path (blue squares) describing the stationary prevalence under small initial density of adopters, which corresponds to diffusion processes from few early adopters to global dissemination. For the ascending path, prevalence first remains zero and then rapidly grows to a high level at a large threshold as $\lambda_1$ or $p$ increases. We call the threshold as the diffusion threshold. The other one is the descending path (red triangles) which depicts the final results under large initial density of adopters. It corresponds to the eradication processes that norms transform from a large initial prevalence to a small one. For the descending path, as $\lambda_1$ or $p$ decreases, prevalence remains a high level and would not recover to zero until it mitigates less than a small threshold. We designate the threshold as the eradication threshold. These two paths form hysteresis loops of stationary prevalence, where the diffusion threshold is much larger than the eradication threshold. This indicates that more efforts of reducing transmissibility are required to make the prevalence recover to zero. Our result explains why it is so difficult to eliminate the rumors and misbeliefs from social networks, especially in the modern media environment where different online social platforms interact.  Besides, noting the dependence of the system's final state on its history, we could alter initial density of adopters to realize the mutual transformation of the two evolutionary routes.

To give a more intuitive illustration about how initial density of adopters affects the final state, in figure  \ref{Time evolutions}, we further present time evolutions of prevalence for four different combinations of transmission parameters. In each subfigure, a single realization indicates  the temporal evolution under a certain value of initial condition ($\rho_0$), which ranges from 0 to 1 and is represented by different colors. The initial conditions show no influence on the prevalence of norms when $\gamma$ is small ($\gamma=0.1$): either go extinct (figure  \ref{Time evolutions}(a)) or outbreak (figure  \ref{Time evolutions}(c) and figure  \ref{Time evolutions}(d)) no matter what $\rho_0$ is. However, figure \ref{Time evolutions}(b) shows completely different phenomena when $\gamma$ is relatively large ($\gamma=0.3$). There appears a threshold of initial density of adopters ($\rho_0=0.09$), below which the prevalence vanishes, while above which the prevalence converges to about 0.3. This indicates that initial density of adopters plays a main role in determining the final state of bistable region. This insight is consistent with empirical studies that critical masses are necessary for initiating social changes, and can not be observed in simple contagion models on multiplex networks \cite{centola2018experimental,xie2011social}.

\begin{figure}
\center
\includegraphics[width=12cm]{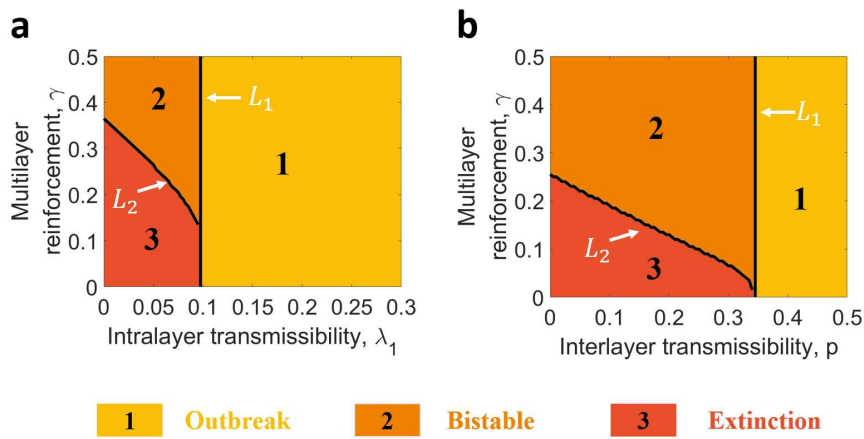}

\caption{Conditions of phase transition for bistability. We provide phase diagrams (a) for the joint effects of multilayer reinforcement and intralayer transmissibility and (b) for the joint effects of multilayer reinforcement and interlayer transmissibility. In each subfigure, the phase plane is composed of three regions: outbreak (region 1), bistable where outbreak and extinction states coexist (region 2), extinction (region 3). The separatrixes $L_1$ and $L_2$ represent the critical parameters when $\rho_0$ is small and when $\rho_0=1$, and are numerically solved by equation (\ref{8}) and equation (\ref{4}). Parameters: (a) $\lambda_2=0.02$, $\mu_1=0.6$, $\mu_2=0.6$, $p=0.1$ and (b) $\lambda_1=0.06$, $\lambda_2=0.04$, $\mu_1=0.6$, $\mu_2=0.6$.}
\label{Phase diagram intralayer}
\end{figure}

Note that the emergence of bistable region is an intriguing physical phenomenon arising from multilayer reinforcement. Here, we further explore the detailed conditions of phase transition for bistability. In figure \ref{Phase diagram intralayer}, we present phase diagram under different combinations of multilayer reinforcement and intralayer transmissibility (figure  \ref{Phase diagram intralayer}(a)) or combinations of multilayer reinforcement and interlayer transmissibility. (figure  \ref{Phase diagram intralayer}(b)). In all subfigures, the phase plane is divided into three regions by two separatrixes ($L_1$ and $L_2$): outbreak, bistable, extinction.  The outbreak state means that the prevalence is positive as long as $\rho_0>0$, and the extinction state represents that the prevalence vanishes no matter what $\rho_0$ is. The bistable state means that norms outbreak for large $\rho_0$ while go extinct for small $\rho_0$. Thus, two separatrixes $L_1$ and $L_2$ are critical values of the parameters under very small $\rho_0$ and $\rho_0=1$, which correspond to the diffusion threshold and the eradication threshold of transmissibility, respectively. $L_1$ and $L_2$ can be numerically solved by equation (\ref{8}) and equation (\ref{4}), respectively.  We find that the solution of equation (\ref{8}) is uncorrelated with multilayer reinforcement ($\gamma$), which is directly reflected by the parallel relationship between $L_1$ and y-axis. Meanwhile, $L_2$ is determined by the joint effects of intralayer transmissibility, interlayer transmissibility and multilayer reinforcement. To sum up, region 1 and region 2 highlight two opportunities for outbreak, which are: (i) to own large transmissibilities and (ii) to own large initial density of adopters with strong multilayer reinforcement. These conclusions can be directly applied to many scenarios such as promoting marketing, designing interventions for rumor spreading and establishing new social norms \cite{karsai2014complex,centola2018experimental}.  We also discuss the impacts of system size on the critical value of multilayer reinforcement that leads to bistability, the diffusion threshold of transmissibility as well as the eradication threshold of transmissibility in \ref{appendix_systemsize}. We find that all the critical values first decrease as $N$ grows when the system is small, and then become stable when $N$ is large ($N>8000$), which also proves the reliability of our large-scale simulations.

\begin{figure}
\center
\includegraphics[width=12cm]{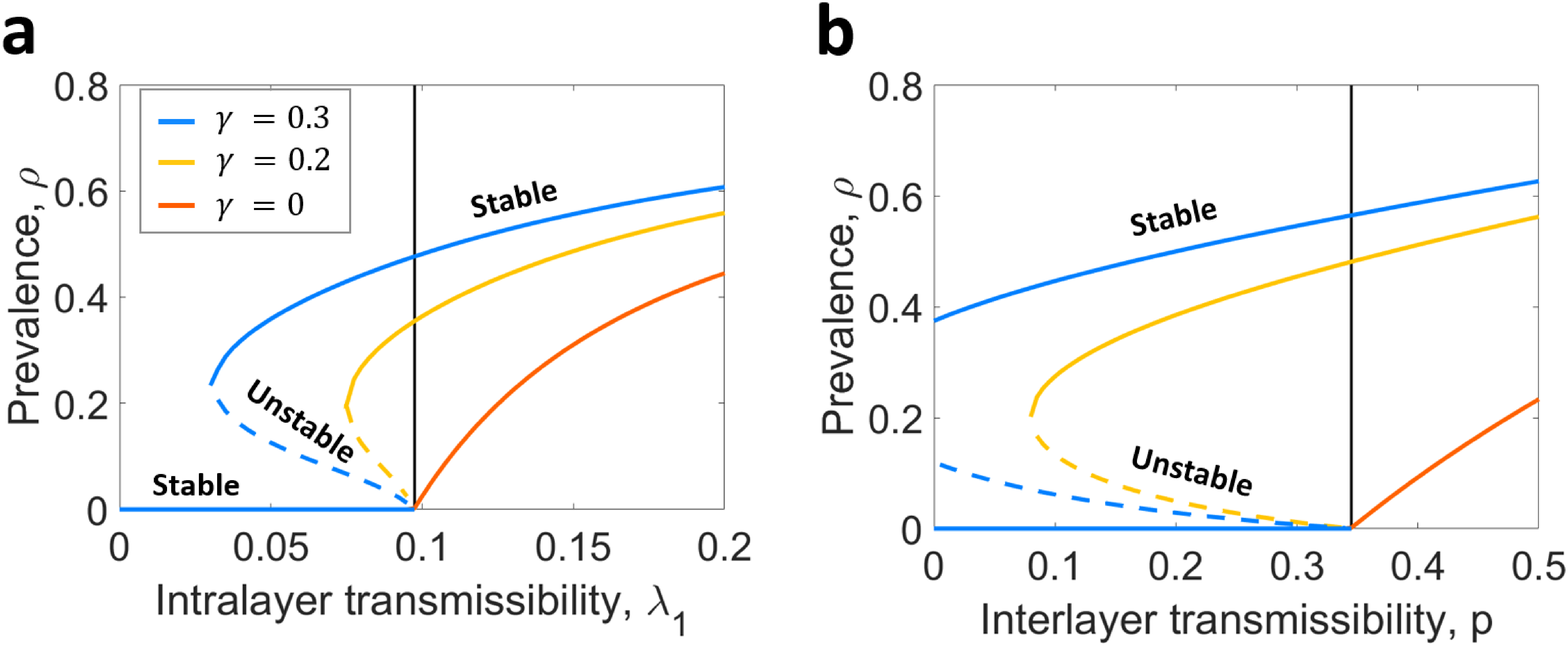}

\caption{Emergence of unstable manifold in bistable region. The prevalence of multiple equilibriums solved by equation (\ref{4}) is shown as a function of (a) intralayer transmissibility and (b) interlayer transmissibility for different multilayer reinforcement. For $\gamma=0.3$ or $\gamma=0.2$ where bistable region exists, the unstable equilibrium manifold (dashed lines) emerges between the stable equilibrium manifolds (solid lines). Vertical line represents the threshold when the system has no multilayer reinforcement. Parameters: (a) $\lambda_2=0.02$, $p=0.1$, $\mu_1=\mu_2=0.6$; (b) $\lambda_1=0.06$, $\lambda_2=0.04$, $\mu_1=\mu_2=0.6$.}
\label{unstable manifold}
\end{figure}

 Furthermore, we show the existence of unstable equilibrium manifold in bistable region in Figure \ref{unstable manifold}. By solving equation (\ref{4}), we present the prevalence of multiple equilibriums with respect to changes in intralayer or interlayer transmissibility. Again, our results illustrate that strong multilayer reinforcement can result in the emergence of discontinuous phase transition and bistability. Moreover, we find that the unstable manifold appears between two stable manifolds in bistable region and the prevalence of unstable equilibriums decreases as $\lambda_1$ or $p$ grows.

\begin{figure}
\center
\includegraphics[width=14cm]{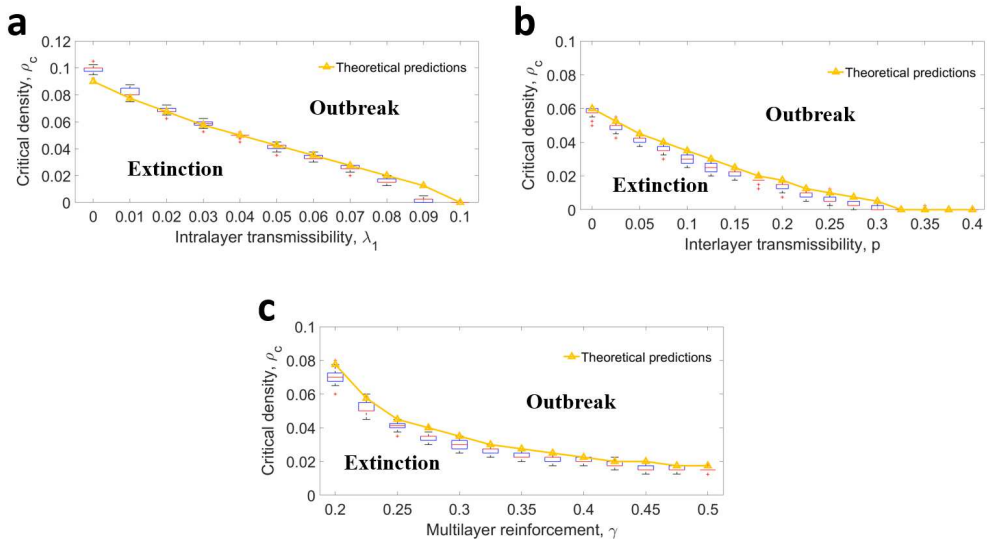}

\caption{Critical density of initial adopters ($\rho_c$) which separates the bistable region into outbreak and extinction. The boxplots present $\rho_c$ as a function of (a) intralayer transmissibility, (b) interlayer transmissibility and (c) multilayer reinforcement. Each box contains 30 independent runs. Theoretical predictions are numerically solved by equation  (\ref{4}) and are represented by yellow triangles. Parameters: $\mu_1=0.6$, $\mu_2=0.6$. In addition, (a) $\lambda_2=0.02$, $p=0.1$, $\gamma=0.4$, (b) $\lambda_1=0.06$, $\lambda_2=0.04$, $\gamma=0.3$, (c) $\lambda_1=0.06$, $\lambda_2=0.04$, $p=0.1$. }
\label{Critical value}
\end{figure}

The bistability phenomenon provides a profound insight that large initial density of adopters could trigger the outbreak even at low values of transmissibility. Here we explore the thresholds of initial conditions ($\rho_c$) under different circumstances, which describe at least how many initial adopters are required for the outbreak of norms in bistable region. In figure  \ref{Critical value}(a)-(c), we use boxplot to present critical value $\rho_c$ as a function of intralayer transmissibility ($\lambda_1$), interlayer transmissibility ($p$) and multilayer reinforcement ($\gamma$), respectively. Each box contains 30 independent simulations. We find that the differences among all simulation results in each box are within 0.02, which implies the stability of $\rho_c$. Besides, the critical density of initial adopters ($\rho_c$) decreases as intralayer transmissibility, interlayer transmissibility or multilayer reinforcement increases. Also, our theoretical solutions well predict simulation results. In particular, we notice that all critical values are small (lower than 0.12), which indicates that manipulating initial condition is a powerful and effective way to control social contagions under the existence of multilayer reinforcement.

\subsection{Complex social contagion on heterogeneous multiplex networks}
\begin{figure}
\center
\includegraphics[width=12cm]{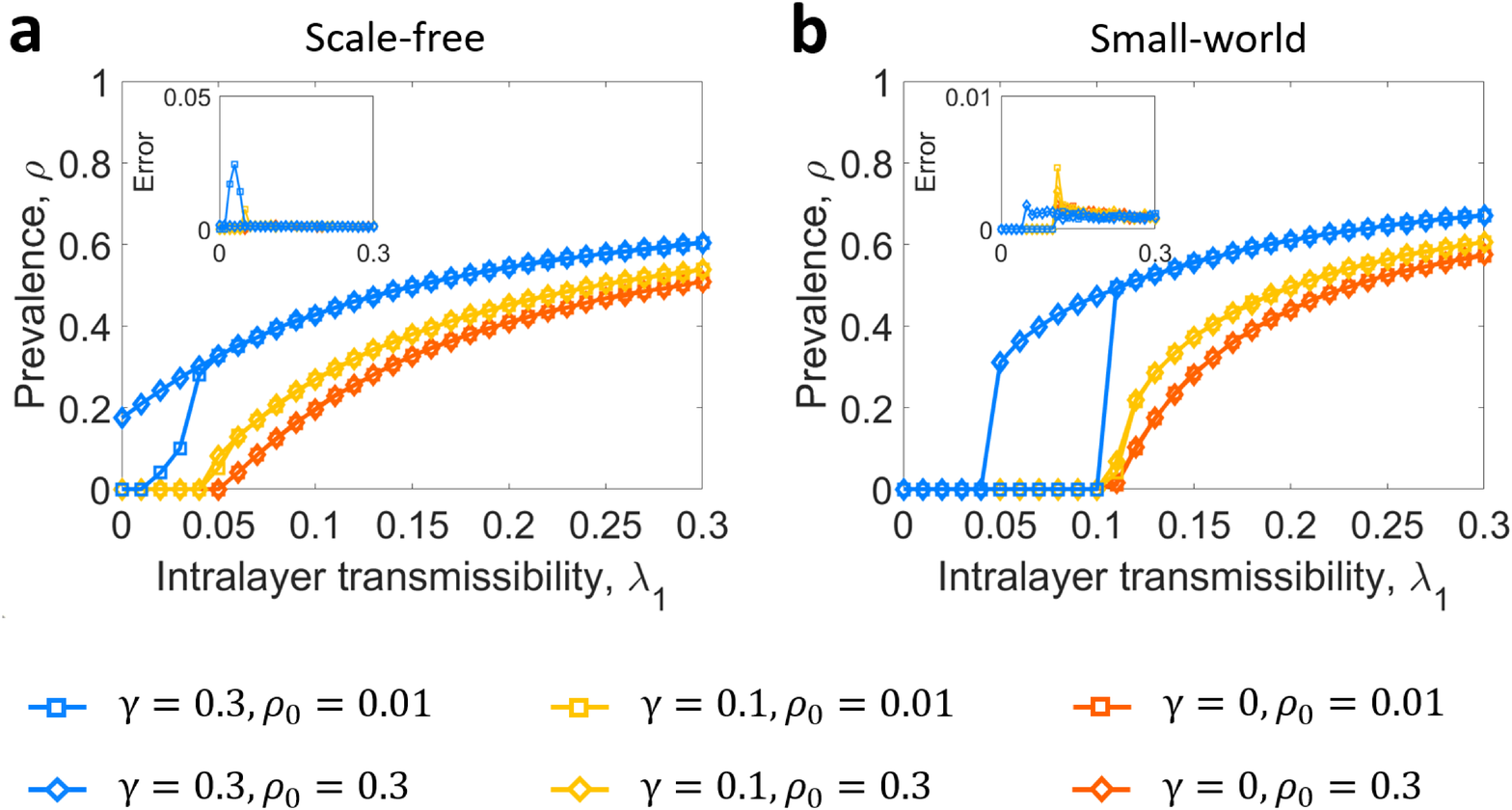}

\caption{Complex social contagion on heterogeneous multiplex networks. The prevalence is presented as a function of intralayer transmissibility for different combinations of multilayer reinforcement and initial density of adopters on (a) SF-SF multiplex networks, exponent parameters of which are $\alpha_1=2.7$ and $\alpha_2=2.9$, and (b) WS-WS multiplex networks with rewiring probability equal to 0.01.  Simulation results are averaged over 30 independent runs, whose standard errors are shown in subplots. Parameters: $\lambda_2=0.02$, $\mu_1=0.6$, $\mu_2=0.6$, $p=0.1$. In addition, $\gamma=0$ (red), $\gamma=0.1$ (yellow), $\gamma=0.3$ (blue) and $\rho_0=0.01$ (square), $\rho_0=0.3$ (diamond).}
\label{SF}
\end{figure}

Real-world networks often have power-law degree distribution, i.e, $p_k\sim k^{-\alpha}$, or small-world characteristics \cite{barabasi1999emergence,watts1998collective}. Studies have revealed that the network structures play an important role in dynamical evolutions \cite{moreno2002epidemic}. Thus, to mimic real situations, in this section, we examine how our model behaves on multiplex systems composed of two scale-free (SF) networks or two Watts-Strogatz small-world (WS) networks, respectively. All networks have 10000 nodes. The rewiring probability of the two WS networks are both 0.01 and the exponent parameters of the two SF networks are $\alpha_1=2.7$ and $\alpha_2=2.9$.

Figure \ref{SF} presents the prevalence as a function of intralayer transmissibility for different values of multilayer reinforcement and initial density of adopters in SF-SF multiplex networks (figure  \ref{SF}(a)) and WS-WS multiplex networks (figure  \ref{SF}(b)). Results show similar phenomena to ER-ER multiplex networks, including promoting effect of multilayer reinforcement and the emergence of bistability when $\gamma$ is large ($\gamma=0.3$).

\begin{figure}
\center
\includegraphics[width=7cm]{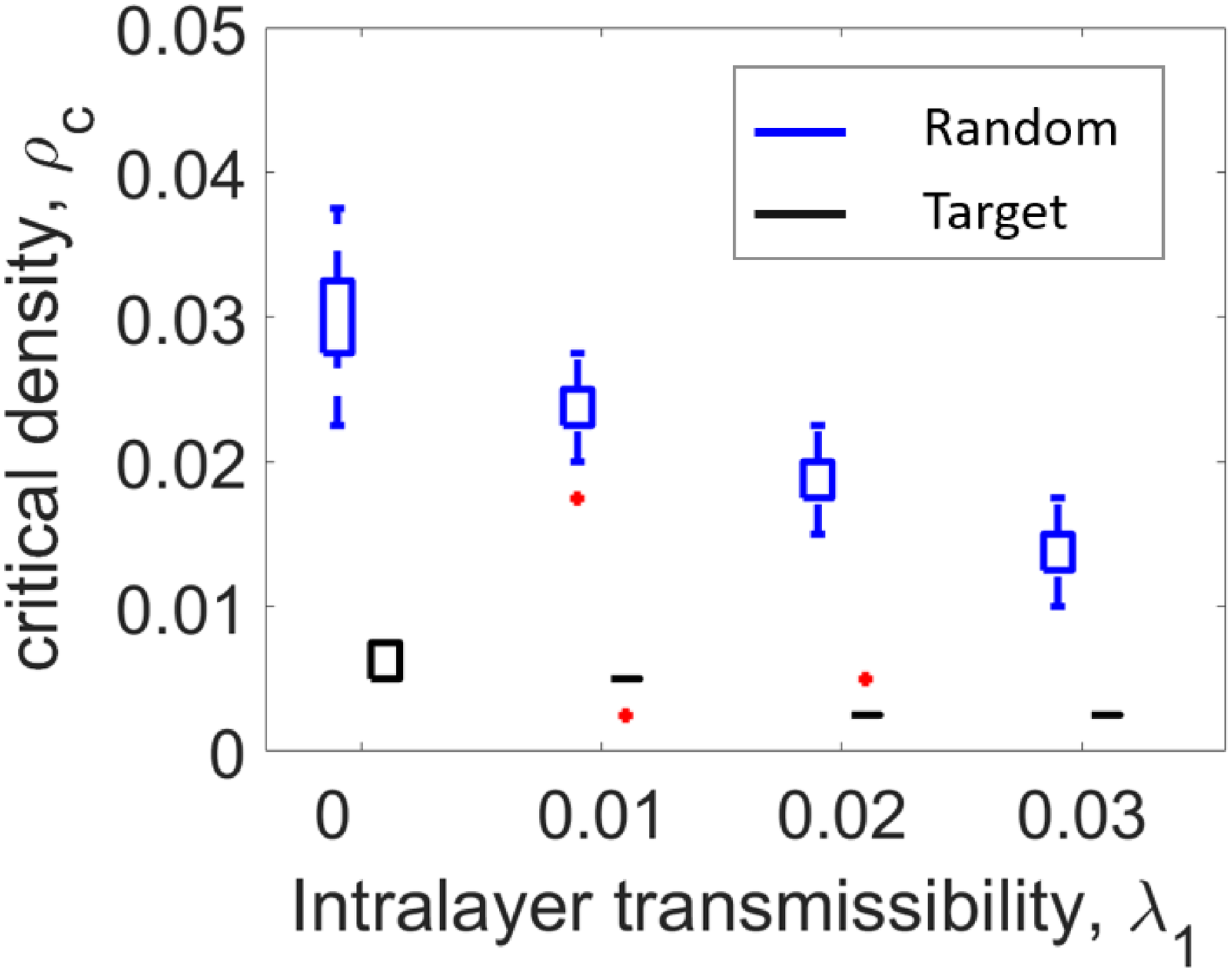}

\caption{Comparison of different strategies for distributing initial seeds. Critical initial density leading to outbreak is presented as a function of intralayer transmissibility when random selection (blue) and target selection (black) are utilized to distribute the initial seeds, respectively. Each box is derived by 30 independent simulations on SF-SF multiplex networks. Parameters: $\lambda_2=0.01$, $p=0.1$, $\gamma=0.3$, $\mu_1=\mu_2=0.6$.}
\label{strategies}
\end{figure}

Finally, we explore the influence of strategies for selecting initial seeds on SF-SF multiplex networks. Two strategies are considered: random selection (randomly selecting initial seeds) and target selection (prefer to select those with higher sum of degree as initial seeds). Figure \ref{strategies} presents the critical density of initial adopters leading to outbreak with respect to changes in intralayer transmissibility when random selection and target selection are adopted, respectively. Results show that a small initial density is enough to result in an outbreak on SF-SF multiplex networks for both strategies. Specifically, the threshold for target selection is much smaller than random selection.

\section{Conclusions and Discussions}\label{conclusions}
Complex social contagion describes complicated behavioral evolutions that cannot be characterized by simple contact-contagion models, such as public opinion formation on controversial events, the spread of conspiracies and the establishment of new social norms \cite{dodds2004universal,lambiotte2019networks}. These risky or polarized collective behaviors are ubiquitous on various social networks and are largely determined by the high-order interactions regarding to the enhancement effects of multiple exposures, i.e., the social reinforcement \cite{lehmann2018complex}. However, it remains unclear how multilayer reinforcement which mimics complex contagion process among multiple social circles influences the final diffusion results.

In this paper, we propose a modified SIS model which incorporates multilayer reinforcement to describe spreading dynamics on multiplex networks. Firstly, we examine how our model behaves on homogeneous multiplex networks. In particular, we stress our efforts on the detailed impact of multilayer reinforcement. A theoretical framework combining pairwise method and mean-field theory is proposed and is verified by large-scale simulations. Interestingly, we find that the multilayer reinforcement induces the emergence of bistability, where extinction and outbreak states coexist. Furthermore, we show that the final state of bistable region is determined by the initial density of adopters whose threshold decreases as intralayer transmissibility, interlayer transmissibility or multilayer reinforcement increases. We also illustrate the hysteresis phenomena where two evolutionary routes occur, known as an ascending path and a descending path, accounting for the phenomenon that rumors are hard to dispel among online social platforms. The detailed conditions of phase transition for bistability are further derived analytically and are shown in phase diagrams. Within the bistable region, we show the existence of unstable manifolds. Results highlight that the chance for the outbreak of social contagion is either to possess large contagion transmissibility or to own large initial density of adopters with strong multilayer reinforcement. Our results are also valid on finite-size heterogeneous multiplex networks.

Our basic yet powerful multilayer reinforcement mechanism reveals the dramatic promoting impacts of complex social contagion on multiplex networks, which indicates the possibility of facilitating spread outbreaks via controlling initial spreaders. These insights are in line with the empirical studies that only large initial density of adopters can successfully initiate social changes \cite{centola2018experimental}. While current results are performed on uncorrelated multiplex networks, further study may extend the framework to partially overlapped multiple networks where edges in different layers are correlated.

\ack{This work is supported by Program of National Natural Science Foundation of China Grant No. 11871004, 11922102, and National Key Research and Development Program of China Grant No. 2018AAA0101100.}

\begin{appendix}
\section{Comparison of Mean-field approximation and simulations}\label{MFA}
\begin{figure}[h!]
\center
\includegraphics[width=12cm]{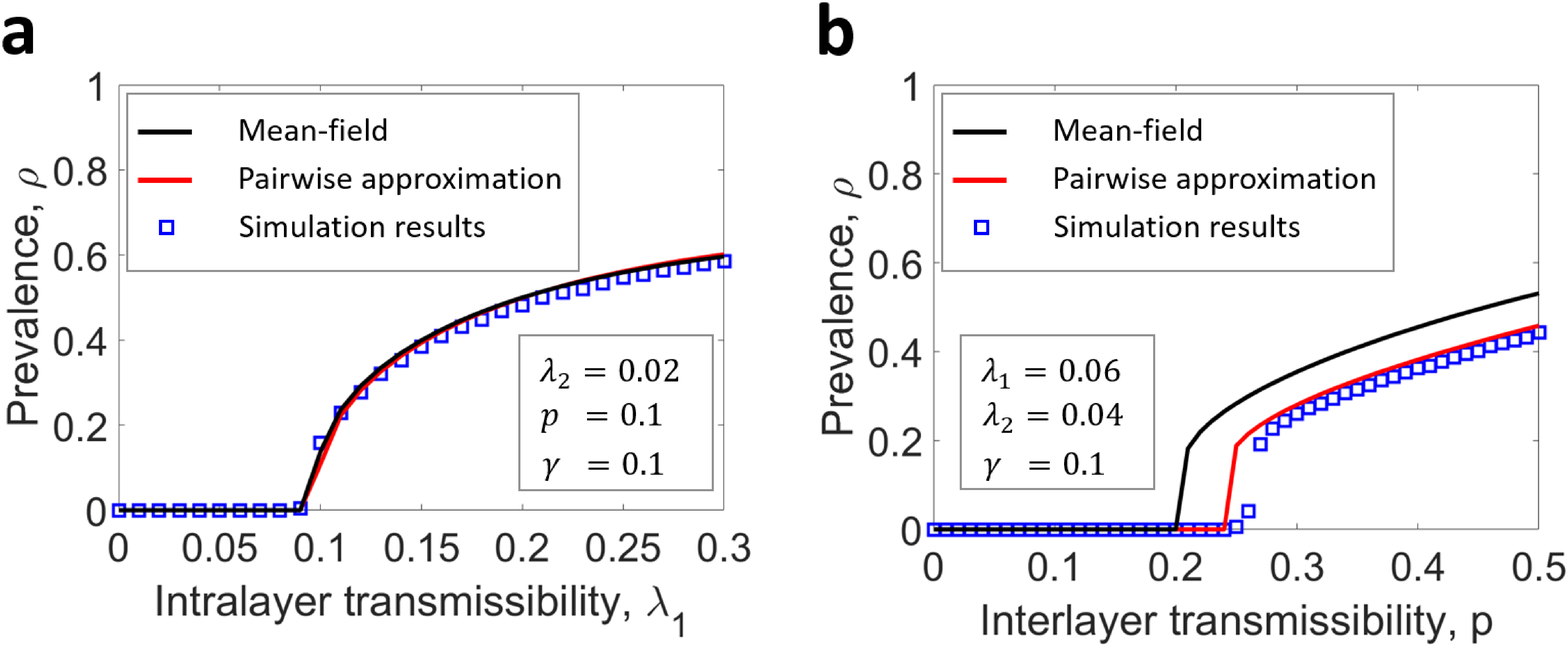}

\caption{Comparison of Mean-field approximation and Pairwise approximation as well as simulations. Prevalence is shown (a) against intralayer transmissibility when interlayer interactions are weak and (b) against interlayer transmissibility. Simulation results (blue squares) are averaged over 30 runs, while mean-field approximation (black line) and pairwise approximation (red line) are solved by equation (\ref{MFE}) and equation (\ref{4}), respectively. Parameters: (a) $\lambda_2=0.02$, $p=0.1$, $\gamma=0.1$, $\mu_1=\mu_2=0.6$; (b) $\lambda_1=0.06$, $\lambda_2=0.04$, $\gamma=0.1$, $\mu_1=\mu_2=0.6$.}
\label{pic_MF}
\end{figure}
Here we consider Mean-field approach to describe the system. Let $\rho_X^j(t),(j=1,2)$ represent the probability that state of node in layer $j$ is $X$. Then we denote $\theta_{MF}^j(t)$ as the probability that ignorant in layer $j$ is not convinced in intralayer contagion process, which can be written as

\begin{equation}
\theta_{MF}^j(t)=(1-\lambda_j\rho_S^j(t))^{\langle k_j\rangle}
\label{A1}
\end{equation}

Next, we denote $\delta_{MF}(t)$ as the probability that ignorant in layer 1 or layer 2 is not affected by multilayer reinforcement.  $\delta_{MF}(t)$ can be calculated by
\begin{equation}
\delta_{MF}(t) = 1 - 0.5*\gamma*(1-(1-\rho_S^1(t))^{\langle k_1\rangle})*(1-(1-\rho_S^2(t))^{\langle k_2\rangle})
\label{A2}
\end{equation}
where $(1-(1-\rho_S^j(t))^{\langle k_j\rangle})$ accounts for the probability that node in layer $j$ has at least one spreader neighbor.

Thus, the evolutionary equations of our dynamical model can be written as
\begin{equation}
\eqalign{
\frac{d\rho_S^1}{dt} &= -\mu_1\rho_S^1 + (1-\rho_S^1)*(1-\theta_{MF}^1(1-p\rho_S^2)\delta_{MF})\\
\frac{d\rho_S^2}{dt} &= -\mu_2\rho_S^2 + (1-\rho_S^2)*(1-\theta_{MF}^2(1-p\rho_S^1)\delta_{MF})\\
}
\label{MFE}
\end{equation}

In figure \ref{pic_MF}, we conduct comparisons of Mean-field approximation and Pairwise approximation as well as simulations. When interlayer interactions are weak, both approximations agree well with simulations (figure \ref{pic_MF}(a)). Nevertheless, Mean-field approximation would deviate from simulation results when interlayer interactions are strong, while Pairwise approximation is still valid (figure \ref{pic_MF}(b)). These indicate that considering dynamical correlations of two layers is necessary for an accurate description of the system, especially under strong interlayer interactions.

\section{complementary studies for figure \ref{intralayer}}\label{appendix}
\begin{figure}[h!]
\center
\includegraphics[width=12cm]{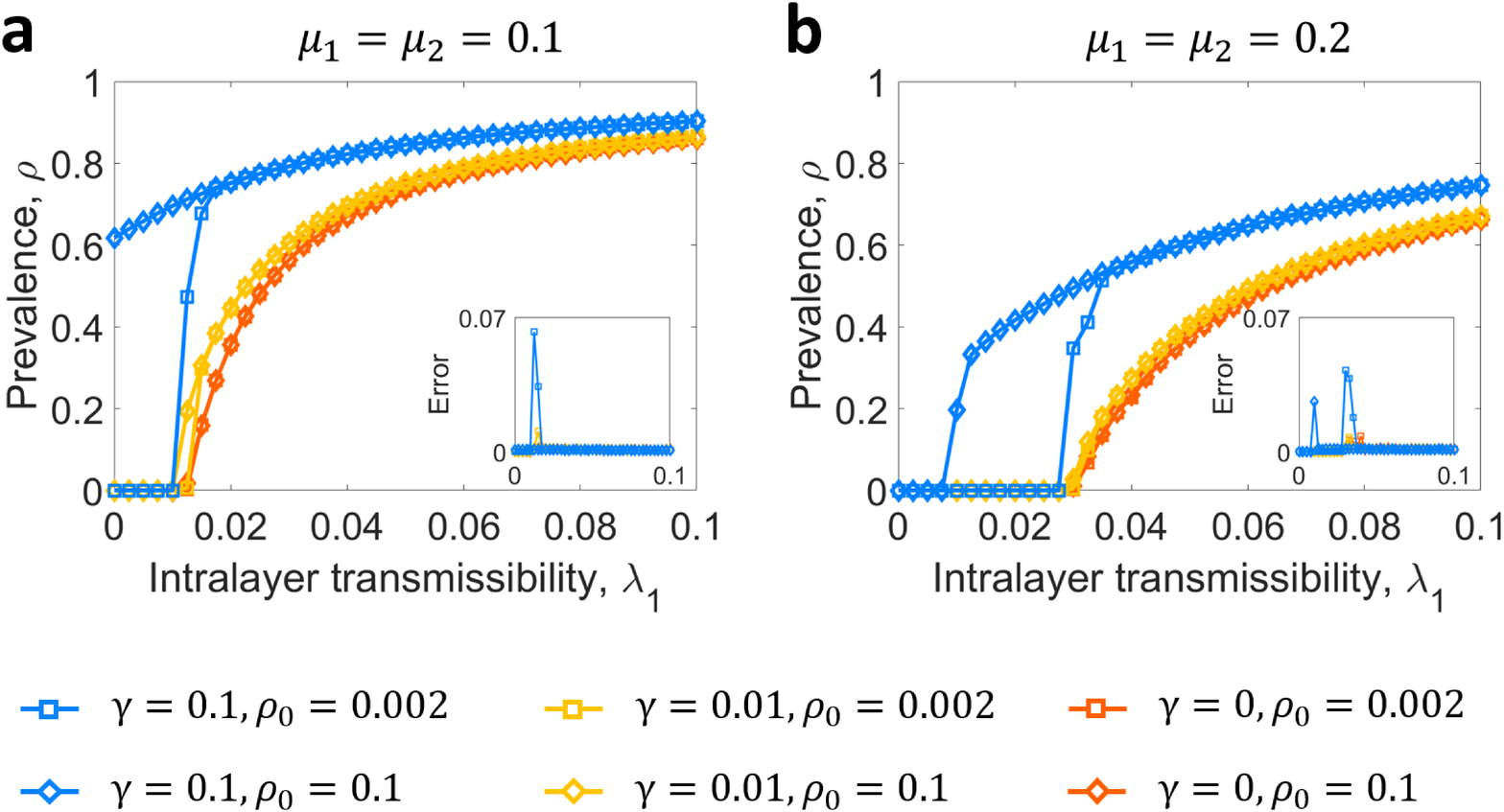}

\caption{The emergence of bistability under different values of recovery. We set (a) $\mu_1=\mu_2=0.1$ and (b) $\mu_1=\mu_2=0.2$, respectively. In each subfigure, the prevalence is shown against intralayer transmissibility under different combinations of multilayer reinforcement and initial density of adopters. Simulation results are averaged over 30 independent runs, whose standard errors are shown in subplots. Parameters: $\lambda_2=0.001$, $p=0.1$. In addition, $\gamma=0$ (red), $\gamma=0.01$ (yellow), $\gamma=0.1$ (blue) and $\rho_0=0.002$ (square), $\rho_0=0.1$ (diamond).}
\label{mu}
\end{figure}
Figure \ref{mu} presents the simulation results for different values of spontaneous recovery. In all subfigures, we observe the emergence of bistable phenomena and discontinuous phase transition when multilayer reinforcement is strong ($\gamma=0.1$). In particular, the case, $\gamma=0.01$, triggers a small bistable region in figure \ref{mu}(a), which indicates that even weak multilayer reinforcement could induce bistability on multiplex networks when the value of spontaneous recovery is small.

\section{Effect of system size}\label{appendix_systemsize}
\begin{figure}[h!]
\center
\includegraphics[width=9.5cm]{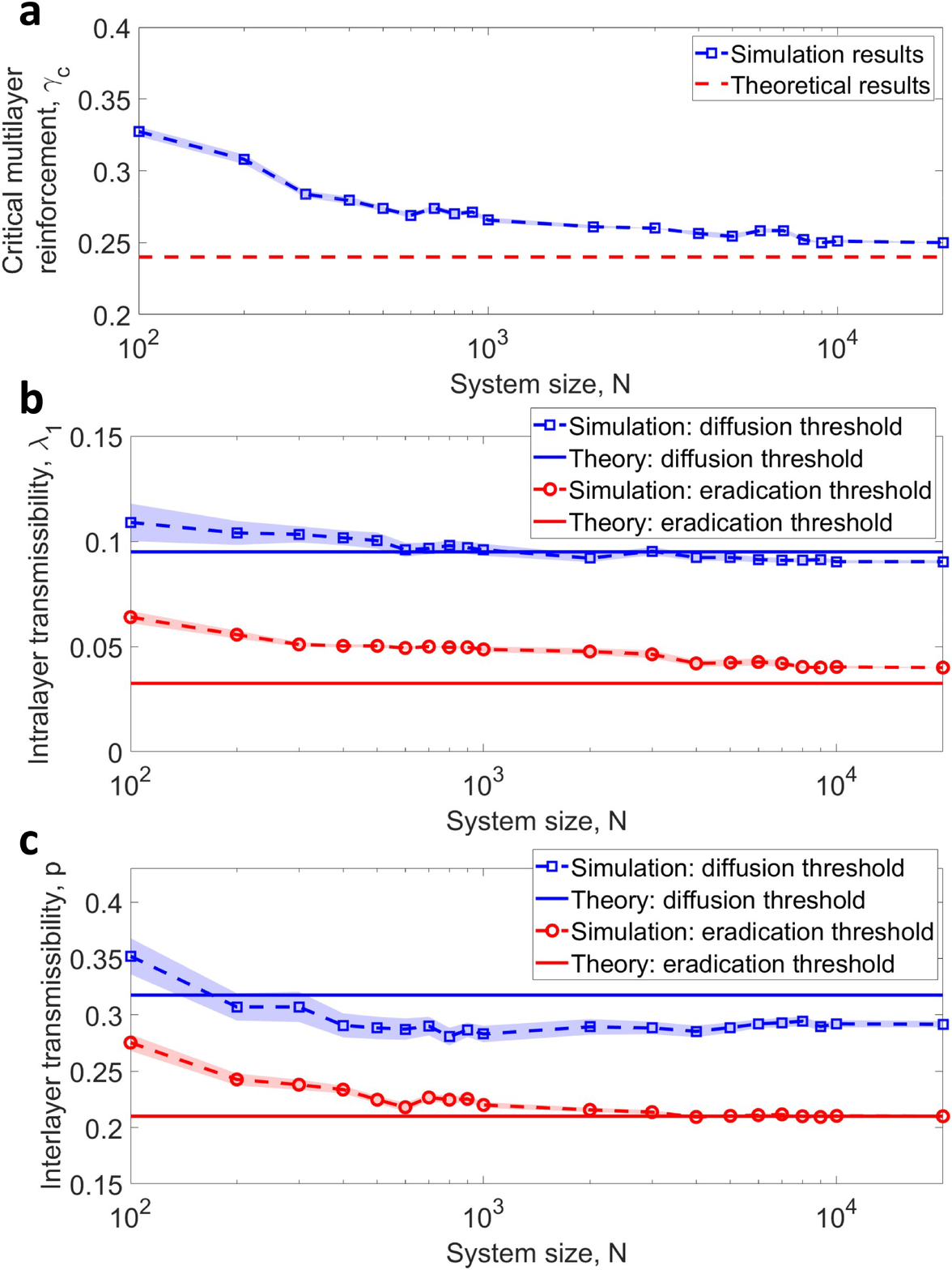}

\caption{Effect of system size. (a) Critical value of multilayer reinforcement ($\gamma_c$) inducing bistability is shown with respect to changes in system size. As system size increases, $\gamma_c$ derived by numerical simulations (blue squares) decreases and finally converges to our theoretical predictions (red line). (b)-(c) Shown are the impacts of system size on the diffusion threshold and the eradication threshold of intralayer transmissibility ($\lambda_1$) and interlayer transmissibility ($p$), respectively. In all subplots, simulation results are averaged over 30 independent runs and light blue/red region represents 95\% confidence interval. Parameters: $\langle k_1\rangle=6$, $\langle k_2\rangle=8$, $\mu_1=\mu_2=0.6$. In addition, (a) $\lambda_1=0.06$, $\lambda_2=0.02$, $p=0.1$; (b) $\lambda_2=0.02$, $p=0.1$, $\gamma=0.3$; (c) $\lambda_1=0.06$, $\lambda_2=0.04$, $\gamma=0.12$.}
\label{network size}
\end{figure}

In this section, we first explore how system size affects critical multilayer reinforcement that induces bistability ($\gamma_c$) in figure \ref{network size}(a). When $N$ is small ($N<8000$), $\gamma_c$ decreases as system size grows. This indicates that small system suppresses the emergence of bistability. When $N$ is large ($N>8000$), $\gamma_c$ is almost unchanged and is in good agreement with our theoretical prediction. It proves the stability of $\gamma_c$ and the validity of our theoretical framework when dealing with large-scale systems.

As for the intralayer or interlayer transmissibility, there arise two different thresholds: the diffusion threshold and the eradication threshold.  The diffusion threshold is the minimum transmissibility ensuring that few early adopters of rumors could result in global diffusion. The eradication threshold is the maximum transmissibility ensuring that the widespread rumors could be finally eliminated. Here we further study how the diffusion threshold and the eradication threshold change by the system sizes (see figure \ref{network size}(b)-(c)). When $N$ is small ($N<8000$), both thresholds decrease as the system size increases. When $N$ is large ($N>8000$), both thresholds are almost unchanged. Meanwhile, our theoretical framework well estimates the diffusion threshold as well as the eradication threshold of intralayer and interlayer transmissibility when $N$ is large.

\end{appendix}

\providecommand{\newblock}{}

\end{document}